\documentclass[aps,prl,twocolumn,groupedaddress]{revtex4}
\usepackage{amssymb}

\usepackage{graphicx}
\usepackage{dsfont}
\usepackage{amsmath}
\usepackage{bm}
\usepackage{times}

\graphicspath{{converted_graphics/}}
\begin{document}

\title{Experimental investigation of dynamical invariants in bipartite entanglement}

\author{O. Jim\'enez Far\'{\i}as\footnote{ Also with Instituto de Ciencias Nucleares, Universidad Nacional Aut\'onoma de M\'exico
(UNAM), Apdo. Postal 70-543, M\'exico 04510 D.F.}}
\affiliation{Instituto de F\'{\i}sica, Universidade Federal do Rio de
Janeiro, Caixa Postal 68528, Rio de Janeiro, RJ 21941-972, Brazil}
\author{A. Vald\'es Hern\'andez}
\affiliation{Instituto de F\'{\i}sica, Universidade Federal do Rio de
Janeiro, Caixa Postal 68528, Rio de Janeiro, RJ 21941-972, Brazil}
\author{G. H. Aguilar}
\affiliation{Instituto de F\'{\i}sica, Universidade Federal do Rio de
Janeiro, Caixa Postal 68528, Rio de Janeiro, RJ 21941-972, Brazil}
\author{P. H. Souto Ribeiro\footnote{Corresponding author e-mail: phsr@if.ufrj.br}}
\affiliation{Instituto de F\'{\i}sica, Universidade Federal do Rio de
Janeiro, Caixa Postal 68528, Rio de Janeiro, RJ 21941-972, Brazil}
\author{S. P. Walborn}
\affiliation{Instituto de F\'{\i}sica, Universidade Federal do Rio de
Janeiro, Caixa Postal 68528, Rio de Janeiro, RJ 21941-972, Brazil}
\author{L. Davidovich}
\affiliation{Instituto de F\'{\i}sica, Universidade Federal do Rio de
Janeiro, Caixa Postal 68528, Rio de Janeiro, RJ 21941-972, Brazil}
\author{Xiao-Feng Qian}
\affiliation{Rochester Theory Center and the Department of Physics \& Astronomy,
University of Rochester, Rochester, New York 14627}
\author{J. H. Eberly}
\affiliation{Rochester Theory Center and the Department of Physics \& Astronomy,
University of Rochester, Rochester, New York 14627}
\date{\today}

\begin{abstract}
The non-conservation of entanglement, when two or more particles interact,
sets it apart from other dynamical quantities like energy and momentum. It
does not allow the interpretation of the subtle dynamics of entanglement as
a flow of this quantity between the constituents of the system. Here we show
that adding a third party to a two-particle system may lead to a conservation
law that relates the quantities characterizing the bipartite entanglement between 
each of the parties and the other two. 
We provide an experimental demonstration of this idea using entangled photons, and 
generalize it to N-partite GHZ states.
\end{abstract}

\maketitle


{\it Introduction.} Proper understanding of the production, quantification and evolution of
quantum entanglement has been a major challenge of quantum physics, with direct implications on  the relevance of this resource for applications in quantum information. A particularly important problem concerns the decay of initially entangled states under the influence of  independent reservoirs acting on each part of the system. While each of these parts undergoes a typical decoherence process, affecting the populations and the coherences of the state, the dynamics of entanglement may differ considerably from local dynamics  \cite{karol, simon, diosi, dodd, dur, yu1, mintert, hein, santos:040305, yu:140403, almeida07, carvalho-2007, salles08, farias09, aolita}.  

Usually, entanglement is not a conserved quantity. For instance, when an initially excited atom decays, releasing a photon into a zero-temperature environment, the initial and final states of the atom-environment system are not entangled, but the atom does get entangled to the environment at intermediate times. We show nevertheless that adding a third party to this two-party system leads to a conservation law involving quantities that measure the bipartite entanglement between each part of the system and the other two parts.  The introduction of this external party unveils therefore a hidden conservation law for entanglement. We demonstrate it experimentally
using twin photons,  which have been useful tools for exploring subtle properties of quantum physics and quantum information \cite{bouwmeester00}. We follow here the strategy of \cite{almeida07,farias09}, which allows one to study the detailed dynamics of entangled states with optical interferometers. We also point out a generalization of this   conservation law to N-partite GHZ states.

{\it Invariants in two-qubit dynamics.} Let $S$ be a qubit and $R$ the reservoir with which it interacts after
\thinspace $t=0.$ The initial product state of the $S$-$R$ system is assumed
to be

\begin{equation}
\rho _{SR}(0)=\rho _{S}(0)\otimes \rho _{R}(0),  \label{rhoinicial}
\end{equation}
with

\begin{equation}
\rho _{S}(0)=\left(
\begin{array}{cc}
\rho _{gg} & \rho _{ge} \\
\rho _{eg} & \rho _{ee}
\end{array}
\right) ,\quad \rho _{R}(0)=\left\vert {\phi _{0}}\right\rangle \left\langle
{\phi _{0}}\right\vert .  \label{rhoSzero}
\end{equation}
The matrix $\rho _{S}$ is written in the basis $\left\{ \left\vert {g}%
\right\rangle ,\left\vert {e}\right\rangle \right\} $ of the ground and
excited states of $S$, and $\left\vert {\phi _{0}}\right\rangle $ stands for
the ground state of the reservoir $R$. At $t=0$ both $S$ and $R$ start to
interact in such a way that the following transformation holds:
\begin{equation}
\begin{array}{l}
\left\vert {g}\right\rangle \left\vert {\phi _{0}}\right\rangle \rightarrow
\left\vert {g}\right\rangle \left\vert {\phi _{0}}\right\rangle , \\
\left\vert {e}\right\rangle \left\vert {\phi _{0}}\right\rangle \rightarrow
\sqrt{1-p}\left\vert {e}\right\rangle \left\vert {\phi _{0}}\right\rangle +%
\sqrt{p}\left\vert {g}\right\rangle \left\vert {\phi }_{1}\right\rangle ,
\end{array}
\label{amplitude}
\end{equation}
where $p=p(t)\in \left[ 0,1\right] $ is a time-dependent parameter such that $%
p(0)=0,$ $p(\infty )=1,$ and $\left\vert {\phi _{1}}\right\rangle $ denotes
the first excited state (orthogonal to $\left\vert {\phi _{0}}\right\rangle $%
) of $R$. The map (\ref{amplitude}) corresponds precisely to the amplitude
damping channel. For different parameterizations $p(t),$ the transformation (%
\ref{amplitude}) represents several physical processes such as the
spontaneous emission of a photon by a two-level atom in a zero-temperature
electromagnetic environment, or the interaction of a two-level atom with a
single mode of the electromagnetic field inside a cavity. According to the
map (\ref{amplitude}), the matrices (\ref{rhoSzero})$\ $evolve into:
\begin{eqnarray}
\rho _{S}(p) &=&\left(
\begin{array}{cc}
1-\rho _{ee}(1-p) & \rho _{ge}\sqrt{1-p} \\
\rho _{eg}\sqrt{1-p} & \rho _{ee}(1-p)
\end{array}
\right) ,\quad  \label{rhop} \\
\rho _{R}(p) &=&\left(
\begin{array}{ccc}
1-\rho _{ee}p & \rho _{ge}\sqrt{p} & ... \\
\rho _{eg}\sqrt{p} & \rho _{ee}p & ... \\
... & ... & ...
\end{array}
\right) ,  \notag
\end{eqnarray}
where \textquotedblleft ...\textquotedblright\ in the expression for $\rho
_{R}(p)$ represents empty rows and columns corresponding to the infinite
remaining null matrix elements. Inspection of the reduced density matrices (%
\ref{rhop}) shows that the information initially contained in the system $S$
is transferred to the system $R$. The transfer is complete at $p=1,$ when
the states of $S$ and $R$ become exchanged.

An important observation regarding the dynamics imposed by the
transformation (\ref{amplitude}) is that the mean number $\langle \hat{N%
}\rangle $ of total excitations,
\begin{equation}
\langle \hat{N}\rangle =\left\langle \hat{n}_{S}(p)+\hat{n}%
_{R}(p)\right\rangle ,  \label{excitation}
\end{equation}
is conserved through the entire evolution. Thus, this restricts the way the
populations (in the referred basis) are transferred. Here $\hat{n}_{S}$ and $%
\hat{n}_{R}$ are the excitation-number operators of the system $S$ and $R$
respectively. For the specific case in which $S$ represents a two level atom
interacting with one mode of the electromagnetic field in a cavity, these
operators are given by $\hat{n}_{S}=\frac{1}{2}(\mathbb{I}-\sigma _{z})$ and
$\hat{n}_{R}=\hat{a}^{\dagger }\hat{a}$. The conservation of $\langle
\hat{N}\rangle $ follows immediately from the expressions for $%
\left\langle \hat{n}_{R}(p)\right\rangle $ and $\left\langle \hat{n}%
_{S}(p)\right\rangle $,
\begin{eqnarray}
\left\langle \hat{n}_{S}(p)\right\rangle &=&\text{Tr}\left[ \rho _{S}(p)\hat{%
n}_{S}\right] =\rho _{ee}(1-p),  \label{medios} \\
\left\langle \hat{n}_{R}(p)\right\rangle &=&\text{Tr}\left[ \rho _{R}(p)\hat{%
n}_{R}\right] =\rho _{ee}p,  \notag
\end{eqnarray}
so that $\langle \hat{N}\rangle =\rho _{ee}.$

In the following, we investigate how the conservation (\ref{excitation})
manifests itself in the evolution  of
the purity $\pi _{i}\equiv $Tr$\rho _{i}^{2}$ of the subsystem $i=S,R$. With
the aid of Eqs. (\ref{rhop}) and (\ref{medios}) it is straightforward to
show that $\pi _{i}(p)$ can be written as
\begin{equation}
\pi _{i}(p)=2\left\langle \hat{n}_{i}(p)\right\rangle ^{2}-2\left\langle
\hat{n}_{i}(p)\right\rangle \Lambda +1,  \label{pi}
\end{equation}
where $\Lambda =1-\left| \rho _{ge}\right| ^{2}/\rho _{ee}.$ Inverting
this expression, we obtain
\begin{equation}
\left\langle \hat{n}_{i}(p)\right\rangle =\tfrac{\Lambda }{2}\pm \tfrac{1}{%
2}\sqrt{\Lambda ^{2}-2(1-\pi _{i}(p))}.  \label{enei}
\end{equation}
Hence we can rewrite the conservation equation (\ref{excitation}) as
\begin{equation}
\rho _{ee}=\Lambda \pm \mathcal{W}_{S}(\Lambda ,p)\pm \mathcal{W}%
_{R}(\Lambda ,p),  \label{conserva2}
\end{equation}
where we have defined $\mathcal{W}_{i}(\Lambda ,p)=\tfrac{1}{2}\sqrt{%
\Lambda ^{2}-2(1-\pi _{i}(p))}.$ As seen in Eq. (\ref{enei}), the
appropriate choice of the sign in front of $\mathcal{W}_{i}$ depends on
whether $\left\langle \hat{n}_{i}(p)\right\rangle $ is greater or smaller
than $\Lambda /2$. If the initial value $\rho _{ee}$ is smaller than or
equal to $\Lambda /2$, then the restriction $\left\langle \hat{n}%
_{i}(p)\right\rangle \leq \rho _{ee}$ implies that $\left\langle \hat{n}%
_{i}(p)\right\rangle \leq \Lambda /2$, and both minus signs should be
taken in (\ref{conserva2}). On the other hand, if $\rho _{ee}$ is larger
than $\Lambda /2$, then the signs in (\ref{conserva2}) depend on the value
of $p.$ From Eqs. (\ref{medios}) it follows that $\left\langle \hat{n}%
_{S}(p)\right\rangle \gtrless $ $\Lambda /2$ whenever $p\lessgtr
1-\Lambda /2\rho _{ee},$ whereas $\left\langle \hat{n}_{R}(p)\right\rangle
\gtrless $ $\Lambda /2$ for every $p\gtrless \Lambda /2\rho _{ee}$. In
both cases the upper/lower inequality sign determine the $\pm $ sign that
should be used.

With the previous results we see that as a consequence of the conservation
of $\langle \hat{N}\rangle ,$ the purities $\pi _{S}(p)$ and $\pi
_{R}(p)$ evolve in such a way that the right hand side of Eq. (\ref
{conserva2}) -- or more generally any function of this argument -- remains
constant during the evolution.


{\it Entanglement conservation.} We recognized that a dynamical invariance property of
the map (\ref{amplitude}) can be used to reveal that more complex
quantities (in this case the purity of each subsystem) can be combined into
new invariant forms. The purity of a system provides a different physical
content from the excitation number, and this is the starting point for the
following analysis. If the quantum state $\rho _{S}$ has purity different
from one, then there exists a larger system in a pure state, so that the
reduced density operator corresponding to $S$ is precisely the mixed state $%
\rho _{S}$. We will refer to the system needed to purify the enlarged system
as $M$. The recognition of this latter system is crucial, not only for the
understanding of the physical origin of states such as $\rho _{S}(0),$ but
also because it throws some light into the global dynamical properties of
(bipartite) entanglement for systems undergoing the interaction modeled by (%
\ref{amplitude}).

Let us suppose that $\rho _{S}(0)$ results from partial tracing over the
system $M$ on the pure general state
\begin{equation}
\left| {\psi (0)}\right\rangle \!=\alpha \left| {M_{1}}\right\rangle
\!\left| {e}\right\rangle +\beta \left| {M_{0}}\right\rangle \!\left| {g}%
\right\rangle +\gamma \left| {M_{1}}\right\rangle \!\left| {g}\right\rangle
+\delta \left| {M_{0}}\right\rangle \!\left| {e}\right\rangle \!,  \label{1}
\end{equation}
where $\left| {M_{0}}\right\rangle ,$ $\left| {M_{1}}\right\rangle $ are two
orthogonal states of $M$. In this case the elements of the initial density
matrix $\rho _{S}(0)$ are $\rho _{ee}=\left| \alpha \right| ^{2}+\left|
\delta \right| ^{2},\rho _{ge}=\beta \delta ^{\ast }+\alpha ^{\ast }\gamma ,$
and $\rho _{gg}=\left| \beta \right| ^{2}+\left| \gamma \right| ^{2}.$ At $%
t=0$ we allow system $S$ to interact with the environment, according to the
transformation (\ref{amplitude}). Then, the initial tripartite state $\left|
{\Psi (0)}\right\rangle =\left| {\psi (0)}\right\rangle \left| {\phi _{0}}%
\right\rangle $ evolves to
\begin{eqnarray}
\left| {\Psi (p)}\right\rangle \!\!\! &=&\!\alpha \left| {M_{1}}%
\right\rangle \!(\sqrt{1-p}\left| {e}\right\rangle \!\left| {\phi _{0}}%
\right\rangle +\!\sqrt{p}\left| {g}\right\rangle \!\left| {\phi }%
_{1}\right\rangle )+  \label{tripartita} \notag \\
&+&\delta \left| {M_{0}}\right\rangle \!(\sqrt{1-p}\left| {e}\right\rangle
\!\left| {\phi _{0}}\right\rangle +\!\sqrt{p}\left| {g}\right\rangle
\!\left| {\phi }_{1}\right\rangle )+  \notag \\ 
&+&(\beta \left| {M_{0}}\right\rangle +\gamma \left| {M_{1}}\right\rangle
)\left| {g}\right\rangle \!\left| {\phi _{0}}\right\rangle .  
\end{eqnarray}
Using (\ref{tripartita}), the density matrix $\rho _{MSR}(p)$ of the
complete system may be constructed, and the reduced density matrices $\rho
_{i}(p)$ with $i=M,R,S$ can be computed. Since $M$ does not interact at all $%
\rho _{M}$ is constant and given by
\begin{equation}
\rho _{M}=\left(
\begin{array}{cc}
\left| \beta \right| ^{2}+\left| \delta \right| ^{2} & \beta \gamma ^{\ast
}+\alpha ^{\ast }\delta \\
\alpha \delta ^{\ast }+\beta ^{\ast }\gamma & \left| \alpha \right|
^{2}+\left| \gamma \right| ^{2}
\end{array}
\right) .  \label{rhoM}
\end{equation}

We return to Eq. (\ref{conserva2}) and observe that, given it was obtained
from a property of the interaction between $R$ and $S$ only, it remains
valid once the system $M$ has been considered. Moreover, as seen from Eq. (%
\ref{tripartita}), the coefficients ($\alpha $ and $\delta $) that determine
$\rho _{ee}$ are precisely the coefficients responsible for the entanglement
between $M$ and the rest of the system. This is an important observation
since it relates $\rho _{ee}$ directly to such entanglement, and hence
allows us to relate $\rho _{ee}$ with the (constant) purity of the system $M$
as follows
\begin{eqnarray}
\rho _{ee} &=&\tfrac{\Lambda }{2}\pm \tfrac{1}{2}\sqrt{\Lambda
^{2}-2(1-\pi _{M})}  \label{rhoee} \\
&=&\tfrac{\Lambda }{2}\pm \mathcal{W}_{M}(\Lambda ).  \notag
\end{eqnarray}
Once more, the $\pm $ sign depends on whether $\rho _{ee}$ is larger (plus
sign) or smaller (minus sign) than $\Lambda /2.$ Introducing this last
expression into Eq. (\ref{conserva2}) leads to
\begin{equation}
\pm \mathcal{W}_{M}(\Lambda )-\tfrac{\Lambda }{2}=\pm \mathcal{W}%
_{S}(\Lambda ,p)\pm \mathcal{W}_{R}(\Lambda ,p).  \label{conservgral}
\end{equation}

As has been stated above, the appropriate choice of signs for each $\mathcal{%
W}_{i}$ is determined by the magnitude of $\rho _{ee}$ relative to $%
\Lambda /2,$ as well as by the value of $p$. An inspection of all the
valid combinations leads finally to the following cases (we omit the
dependence on $\Lambda $) :
\begin{equation}
\tfrac{\Lambda }{2}+\mathcal{W}_{M}=\mathcal{W}_{S}(p)+\mathcal{W}%
_{R}(p),\quad p\in \lbrack 0,1],  \label{case1}
\end{equation}
whenever $\rho _{ee}\leq \frac{\Lambda }{2},$ and
\begin{equation}
\tfrac{\Lambda }{2}-\mathcal{W}_{M}=\!\!\left\{ \!\!
\begin{array}{cl}
\mathcal{W}_{R}(p)-\mathcal{W}_{S}(p), & p<1-\frac{\Lambda }{2\rho _{ee}}
\\
\mathcal{W}_{S}(p)+\mathcal{W}_{R}(p), & p\in \lbrack 1-\frac{\Lambda }{%
2\rho _{ee}},\frac{\Lambda }{2\rho _{ee}}] \\
\mathcal{W}_{S}(p)-\mathcal{W}_{R}(p), & p>\frac{\Lambda }{2\rho _{ee}},
\end{array}
\right.  \label{cases2-4}
\end{equation}
whenever $\rho _{ee}>\frac{\Lambda }{2}$.

Since the tripartite state remains pure during the evolution, each $\pi _{i}$
may be regarded as a quantitative measure of bipartite entanglement between
system $i$ and the rest. Alternatively, we may use a measure based on the
Schmidt weight $K_{i} $ \cite{Qian-Eberly}, which are related to $\pi_{i}$ according to $%
\pi _{i}(p)=K_{i}^{-1}(p)$. As a result, Eqs. (\ref{case1})-(\ref{cases2-4})
stand as conservation relations for bipartite entanglement. We conclude
that, once the original $S$-$R$ system is enlarged to include system $M$
needed for purification, the conservation law (\ref{excitation}) acquires a
new significance in terms of conservation of bipartite entanglement. The
initial entanglement between $S$ and $M$ turns into entanglement between $M$
and $R$, and the entanglement transfer is described by Eqs. (\ref{case1})-(%
\ref{cases2-4}).

In the particular case in which the initial density matrix $\rho _{S}(0)$ is
diagonal so that $\Lambda =1$ (for example if $\delta =\gamma =0$ in Eq. (%
\ref{1})), then the conservation relations (\ref{case1})-(\ref{cases2-4})
involve the quantities $W_{i}(p)\equiv \mathcal{W}_{i}(1,p)=\sqrt{\tfrac{1}{%
2K_{i}(p)}-\tfrac{1}{4}}$, which are explicitly written as
\begin{eqnarray}
W_{S}(p) &=&|\rho _{ee}(1-p)-\tfrac{1}{2}|,  \notag \\
W_{R}(p) &=&|\rho _{ee}p-\tfrac{1}{2}|,  \label{Wfunctions} \\
W_{M} &=&|\rho _{ee}-\tfrac{1}{2}|,  \notag
\end{eqnarray}
as follows from the definition of $W_{i}$ and Eqs. (\ref{medios})-(\ref{pi}).


The conservation relations (\ref{case1})-(\ref{cases2-4}) for $\Lambda =1 $
hold even when we consider system $M$ to be in an initial state entangled
with \thinspace $N$ qubits $S_{j}$ ($j=1,2,...N$)$.$ This allows us to find
new global quantities that are conserved during the evolution. For example, the expression equivalent to Eq.~(\ref{case1}) for the $N+1$ GHZ-type state
\begin{equation}
\left\vert {\chi _{N}}\right\rangle =\alpha \left\vert {M_{1}}\right\rangle
\Pi _{j}^{N}\left\vert {e_{j}}\right\rangle +\beta \left\vert {M_{0}}%
\right\rangle \Pi _{j}^{N}\left\vert {g_{j}}\right\rangle  \label{ghz}
\end{equation}
is:
\begin{equation}
W_{M}+\tfrac{1}{2}=\frac{1}{N}\sum_{j=1}^{N}\left[
W_{S_{j}}(p_{j})+W_{R_{j}}(p_{j})\right] .  \label{geral}
\end{equation}



\begin{figure}[h] 
  \centering
  \includegraphics[bb=2 0 504 503,width=6cm,height=6cm,keepaspectratio]{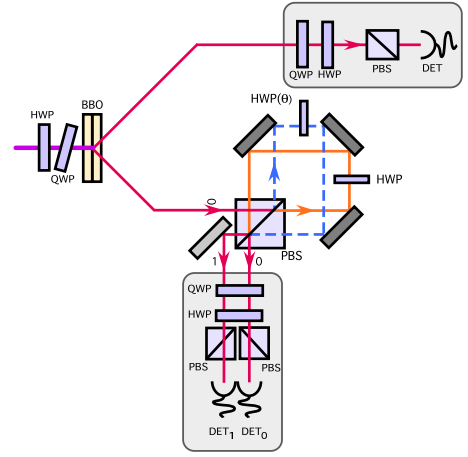}
  \caption{Experimental setup.}
  \label{fig:expsetup}
\end{figure}

{\it Experimental realization.} We verified the validity of Eqs.~(\ref{case1}) and (\ref{cases2-4}) using
polarization-entangled photons generated from spontaneous parametric down
conversion (SPDC). The polarization entanglement is prepared with a
two-crystal source \cite{kwiat99}. The experimental setup is shown in figure
\ref{fig:expsetup}. A 325nm cw He-Cd laser is used to pump two 1mm long
type-I BBO crystals. The down-converted photons are spectrally filtered with 
10nm bandwidth interference filters, and spatially filtered through 2mm detection apertures,
before detection with single-photon counting modules. Identifying
the polarization state of photon 2 as the system $S$ and the polarization state of
photon 1 as the third party $M$, the source is set up to produce the
initial state
\begin{equation*}
\left| {\Phi }\right\rangle =\alpha \left| {V}\right\rangle _{M}\left| {V}%
\right\rangle _{S}+\beta \left| {H}\right\rangle _{M}\left| {H}\right\rangle
_{S}.
\end{equation*}
Let us further identify the longitudinal spatial mode of photon 2 as the
reservoir $R$, and call the initial spatial mode $\left| {0}\right\rangle $.
A displaced Sagnac interferometer with a nested wave-plate can be used to
implement the following transformation on the polarization and spatial mode
of photon 1 \cite{almeida07,salles08,farias09}
\begin{equation}
\begin{array}{l}
\left| {H}\right\rangle \left| {0}\right\rangle \rightarrow \left| {H}%
\right\rangle \left| {0}\right\rangle , \\
\left| {V}\right\rangle \left| {0}\right\rangle \rightarrow \cos \theta
\left| {V}\right\rangle \left| {0}\right\rangle +\sin \theta \left| {H}%
\right\rangle \left| {1}\right\rangle ,
\end{array}
\label{amplitudex}
\end{equation}
where $\left| {0}\right\rangle $ and $\left| {1}\right\rangle $ refer to
different spatial modes and $\theta $ is twice the angle of the half-wave
plate. It has been demonstrated with quantum process tomography that this
interferometer implements the amplitude damping channel with fidelities as
high as $\sim 0.95$ \cite{farias09}. Identifying the polarization states $%
\{\left| {H}\right\rangle ,\left| {V}\right\rangle \}$ with the
system states $\{\left| {g}\right\rangle ,\left| {e}\right\rangle \}$, the
spatial modes $\{\left| {0}\right\rangle ,\left| {1}\right\rangle \}$ with
the reservoir states $\{\left| {\phi _{0}}\right\rangle ,\left| {\phi _{1}}%
\right\rangle \}$, and $p=\sin ^{2}\theta $, the transformation (\ref
{amplitudex}) becomes equivalent to (\ref{amplitude}), provided $0\leq
\theta \leq \pi /2$. The initial state $\left| {\Psi }\right\rangle
_{SM}\left| {0}\right\rangle _{R}$ evolves to
\begin{eqnarray}
\left| {\Phi (\theta )}\right\rangle  &=& \beta \left| {H}\right\rangle
_{M}\left| {H}\right\rangle _{S}+\alpha \cos \theta \left| {V}\right\rangle
_{M}\left| {V}\right\rangle _{S})\left| {0}\right\rangle _{R}  \notag \\
&&+\alpha \sin \theta \left| {V}\right\rangle _{M}\left| {H}\right\rangle
_{S}\left| {1}\right\rangle _{R},  \label{phievol}
\end{eqnarray}
after propagation through the interferometer. The final state is equivalent
to $\left| {\Psi (p)}\right\rangle $ given in equation (\ref{tripartita})
with $\gamma =\delta =0$. It is important to note that partially tracing
over any two of the three subsystems of the state (\ref{phievol}) leads to a
diagonal reduced density matrix. Then, the purities $\pi _{j}$ (or
equivalently, the Schmidt weights $K_{j}$) and each $W_{j}$ ($j=S,M,R$) term
in Eq.~(\ref{case1}) can be determined by local population measurements,
made with the detectors 0 and 1 shown in Fig.~(\ref{fig:expsetup}).

By rotating the half-wave plate (HWP) in the pump beam, we selected
different values of $\alpha $ and $\beta $. For instance, we selected $%
\left| \alpha \right| ^{2}=\rho _{ee}=0.31,0.5,0.73$ with corresponding
purities 0.97, 0.94 and 0.96, so that the conservation laws could be applied
directly. These purities were calculated from quantum state tomography of
the initial states (after passage through the interferometer with $\theta =0$%
, see below). Imperfect purity is probably due to spatial walk-off in the
crystal and imperfect alignment of the interferometer. These
parameters indicate that the experimental state is quite close the the ideal
initial pure state.

\begin{figure}[t] 
  \centering
  \includegraphics[bb=99 177 497 600,width=10cm,height=6cm,keepaspectratio]{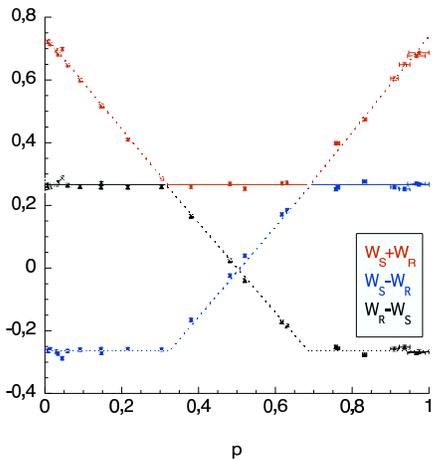}
  \caption{Experimental results for $\protect\rho _{ee}=0.73$, $\protect%
\Lambda =1$. The coloured lines correspond to the functions $%
W_{R}(p)-W_{S}(p)$ (black), $W_{S}(p)+W_{R}(p)$ (red) and $W_{S}(p)-W_{R}(p)$
(blue). The continuous line represents the invariant $I_{SR}$.}
  \label{invarianta}
\end{figure}

\begin{figure}[b] 
  \centering
  \includegraphics[bb=70 177 500 600,width=10cm,height=6cm,keepaspectratio]{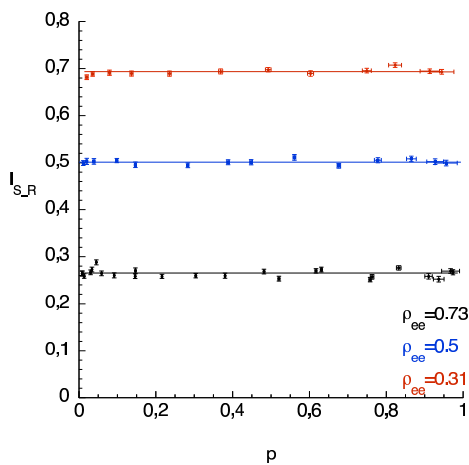}
  \caption{The invariant $I_{SR}$ for different values of $\protect\rho _{ee}$} 
  \label{invariantes}
\end{figure}

Projective measurements on $S$ and $M$ are performed using wave-plates and
polarizing beam splitters to project onto polarization states, while
projection onto spatial modes is performed by placing detectors in mode $0$
or $1$. We performed measurements for several values of $p=\sin ^{2}\theta $
($0\leq \theta \leq \pi /2$) characterizing the amplitude damping channel in
(\ref{amplitudex}).

Figure \ref{invarianta} shows the theoretical curves and the experimental
data of each of the three functions that define the invariant $I_{SR}=\tfrac{%
\Lambda }{2}-\mathcal{W}_{M},$ according to Eq. (\ref{cases2-4}) for the
case $\rho _{ee}=0.73$ and $\Lambda =1$. The black curve corresponds to $%
W_{R}(p)-W_{S}(p),$ the red one to $W_{S}(p)+W_{R}(p),$ and the blue one to $%
W_{S}(p)-W_{R}(p).$ The invariant (piecewise) function $I_{SR}$ is here
stressed by the continuous line and corresponds, as follows from Eq. (\ref
{cases2-4}) (with $\Lambda =1$), to one of the above curves depending on
whether $p$ is in the interval $\left[ 0,1-\frac{1}{2\rho _{ee}}\right) ,%
\left[ 1-\frac{1}{2\rho _{ee}},\frac{1}{2\rho _{ee}}\right] $ or $\left(
\frac{1}{2\rho _{ee}},1\right] .$

Figure \ref{invariantes} shows both the theoretical and experimental curves
of the invariant quantity $I_{SR}$ for three different values of $\rho _{ee}$
(in all cases $\Lambda =1$). The red and blue curves, corresponding
respectively to $\rho _{ee}=0.31$ and $\rho _{ee}=0.5$, represent the
invariant sum $I_{SR}=\tfrac{1}{2}+W_{M}=W_{S}(p)+W_{R}(p)$ (see Eq. (\ref
{case1})), whereas the black curve, corresponding to $\rho _{ee}=0.73,$
represents the invariant $I_{SR}=\tfrac{1}{2}-W_{M}$ given by the expression
(\ref{cases2-4}). From figures \ref{invarianta} and \ref{invariantes} we see
that the experimental data fit the theoretical curves within the precision
of the measurements, thus demonstrating experimentally the conservation
relations (\ref{case1})-(\ref{cases2-4}).

In conclusion, we showed that by adding a third party in the interaction
between two two-level systems, it is possible to obtain quantities that
are invariants in the evolution of the entanglement. We performed an experimental
demonstration of the process using entangled photons and verified the invariance
of these quantitites.   

The authors acknowledge the Brazillian funding agencies CNPq, CAPES and FAPERJ.
This work is part of the National Institute for Science and Tecnology for Quantum 
Information, and it was supported in the US by DARPA HR0011-09-1-0008 and ARO W911NF-09-1-0385.
A.V.H. and O.J.F were funded by the Consejo Nacional de Ciencia y Tecnolog\'{\i}a, M\'exico.


\begin{thebibliography}{10}

\bibitem{karol}
K.~\ifmmode~\dot{Z}\else \.{Z}\fi{}yczkowski, P.~Horodecki, M.~Horodecki, and
  R.~Horodecki,
\newblock Phys. Rev. A {\bf 65}, 012101 (2001).

\bibitem{simon}
C.~Simon and J.~Kempe,
\newblock Phys. Rev. A {\bf 65}, 052327 (2002).

\bibitem{diosi}
L.~Di\'osi,
\newblock Progressive decoherence and total environmental disentanglement,
\newblock in {\em Irreversible Quantum Dynamics}, edited by F.~Benatti and
  R.~Floreanini, pp. 157--163, Springer, Berlin, 2003.

\bibitem{dodd}
P.~J. Dodd and J.~J. Halliwell,
\newblock Phys. Rev. A {\bf 69}, 052105 (2004).

\bibitem{dur}
W.~D\"ur and H.-J. Briegel,
\newblock Phys. Rev. Lett. {\bf 92}, 180403 (2004).

\bibitem{yu1}
T.~Yu and J.~H. Eberly,
\newblock Phys. Rev. Lett. {\bf 93}, 140404 (2004).


\bibitem{mintert}
F.~Mintert, A.~R.~R. Carvalho, M.~Ku\'s, and A.~Buchleitner,
\newblock Phys. Rep. {\bf 415}, 207 (2005).

\bibitem{hein}
M.~Hein, W.~D\"ur, and H.-J. Briegel,
\newblock Phys. Rev. A {\bf 71}, 032350 (2005).


\bibitem{santos:040305}
M.~F. Santos, P.~Milman, L.~Davidovich, and N.~Zagury,
\newblock Phys. Rev. A {\bf 73}, 040305 (2006).

\bibitem{yu:140403}
T.~Yu and J.~H. Eberly,
\newblock Phys. Rev. Lett. {\bf 97}, 140403 (2006), Science {\bf 323}, 598 (2009).

\bibitem{almeida07}
M.~P. Almeida {\em et~al.},
\newblock Science {\bf 316}, 579 (2007).

\bibitem{carvalho-2007}
A.~R.~R. Carvalho, F.~Mintert, S.~Palzer, and A.~Buchleitner,
\newblock Eur. Phys. J. D {\bf 41}, 425 (2007).


\bibitem{salles08}
A.~Salles {\em et~al.},
\newblock Phys. Rev. A {\bf 78}, 022322 (2008).

\bibitem{farias09}
O.~J. Far{\'{i}}as, C.~L. Latune, S.~P. Walborn, L.~Davidovich, and P.~H.~S.
  Ribeiro,
\newblock Science {\bf 324}, 1414 (2009).

\bibitem{aolita}
M.~L. Aolita, R.~Chaves, D.~Cavalcanti, A.~Ac\'{i}n, and L.~Davidovich,
\newblock Phys. Rev. Lett. {\bf 100}, 080501 (2008).

\bibitem{bouwmeester00}
D.~Bouwmeester, A.~Ekert, and A.~Zeilinger, editors,
\newblock {\em The Physics of Quantum Information} (Springer Verlag, Berlin,
  2000).

\bibitem{Qian-Eberly}
See, e.g., X.F. Qian and J.H. Eberly, for a discussion in arXiv:1009.5622 (2010).

\bibitem{kwiat99}
P.~G. Kwiat, E.~Waks, A.~G. White, I.~Appelbaum, and P.~H. Eberhard,
\newblock Phys. Rev. A. {\bf 60}, R773 (1999).

\end{thebibliography}
\end{document}